\documentclass{rspublic}

\usepackage{graphicx}
\usepackage{times}
\usepackage{latexsym}
\usepackage{amssymb}

\newcommand{\be}{\begin{equation}}
\newcommand{\ee}{\end{equation}}

\newcommand{\Sp}{\,\,\,\,\,\,}

\newcommand{\one}{\mbox{$1 \hspace{-1.0mm}  {\bf l}$}}

\newcommand{\sx}{\sigma^x}
\newcommand{\sy}{\sigma^y}
\newcommand{\sz}{\sigma^z}
\newcommand{\ket}[1]{\left | #1\right \rangle}
\newcommand{\bra}[1]{\left \langle #1\right |}

\begin{document}

\title[Geometric phases and criticality in spin systems]
{Geometric phases and criticality in spin systems}

\author[Jiannis K. Pachos and Angelo C. M. Carollo]
{Jiannis K. Pachos and Angelo C. M. Carollo}

\affiliation{Department of Applied Mathematics and Theoretical Physics, \\
University of Cambridge, Wilberforce Road, Cambridge CB3 0WA, UK}

\label{firstpage}

\maketitle

\begin{abstract}{Berry phases, critical phenomena, XY model}

A general formalism of the relation between geometric phases produced by
circularly evolving interacting spin systems and their criticality behavior
is presented. This opens up the way for the use of geometric phases as a
tool to study regions of criticality without having to undergo a quantum
phase transition. As a concrete example a spin-1/2 chain with XY
interactions is presented and the corresponding geometric phases are
analyzed. The generalization of these results to the case of an arbitrary
spin system provides an explanation for the existence of such a relation.

\end{abstract}

\section{Introduction}

A few conceptual advances in quantum physics have been as exciting or as
broadly studied as geometric phases. They lie within the heart of quantum
mechanics giving a surprising connection between the geometric properties of
evolutions and their dynamics. Historically, the first such effect was
presented by Aharonov \& Bohm (1959), where the quantum state of a charged
particle acquires a phase factor when it moves along a closed path in the
presence of a magnetic field. Since then a series of similar effects have
been considered and experimentally verified. A more intriguing case where a
quantum state is changed in a circular fashion was studied by Berry (1984)
which led to the generation of the celebrated Berry phase. This effect does
not need the presence of electromagnetic interactions and due to its
abstract and general nature has found many applications~(Shapere {\it et
al.} 1989; Bohm {\it et al.} 2003).

A characteristic that all non-trivial geometric evolutions have in common is
the presence of non-analytic points in the energy spectrum. At these points
the state of the system is not well defined due to their degenerate nature.
One could say that the generation of a geometric phase is a witness of such
singular points. Indeed, the presence of degeneracy at some point is
accompanied by curvature in its immediate neighborhood and a state that is
evolved along a closed path is able to detect it. These points or regions,
are of great interest to condensed matter or molecular physicists as they
determine, to a large degree, the behavior of complex quantum systems. The
geometric phases are already used in molecular physics to probe the presence
of degeneracy in the electronic spectrum of complex molecules. Initial
considerations by Herzberg and Longuet-Higgins (1963), revealed a sign
reversal when a real Hamiltonian is continuously transported around a
degenerate point. Its generalization to the complex case was derived by
Stone (1976) and an optimization of the real Hamiltonian case was performed
by Johansson \& Sj\"oqvist (2004).

Geometric phases have been associated with a variety of condensed matter and
solid state phenomena (Thouless {\it et al.} 1982; Resta 1994; Nakamura \&
Todo 2002; Ryu \& Hatsugai 2006). Nevertheless, their connection to quantum
phase transitions has only been shown recently by Carollo \& Pachos (2005).
It was farther elaborated by Zhu (2005), where the critical exponents were
evaluate from the scaling behavior of geometric phases, and by Hamma (2006),
who showed that geometric phases can be used as a topological test to reveal
quantum phase transitions. In essence, quantum phase transitions describe
the abrupt changes on the macroscopic behavior of a system resulting from
small variations of external parameters. These critical changes are caused
by the presence of degeneracies in the energy spectrum and are characterized
by long range quantum correlations. This is an exciting area of research
that considers a variety of effects such as the quantum Hall effect and high
$T_c$ superconductivity.

Here we exploit geometric phases as a tool to probe quantum phase
transitions in many-body systems. This provides the means to detect, not
only theoretically, but also experimentally the presence of criticalities.
Apart from the academic interest this approach may have certain advantages.
In particular, the geometric evolutions do not take the system through a
quantum phase transition. The latter is hard to physically implement as it
is accompanied by multiple degeneracies that can take the system away from
its ground state. Hence, they provide a way to probe criticalities in a
physically appealing way. Moreover, the geometric phases provide a non-local
object that might be useful to probe critical phenomena that are
undetectable by local order parameters. The latter consist an exciting field
of current research (Wen 2002).

As an explicit example we employ the one dimensional XY model in the
presence of a magnetic field. This model is analytically solvable and it
offers enough control parameters to support geometric evolutions. By
explicit calculations we observe that an excitation of the model obtains a
non-trivial geometric phase if and only if it circulates a region of
criticality. The generation of this phase can be traced down to the presence
of degeneracy of the energy at the critical point in a similar way used in
molecular systems. The geometric phase can be used to extract information on
the critical exponents that completely characterize the critical behavior. A
generalization of the results to the case of an arbitrary spin system is
demonstrated. Finally, a physical implementation of the XY model and their
corresponding geometric evolutions is proposed with ultra-cold atoms
superposed by optical lattices~(Pachos \& Rico 2005). The independence of
the generated phase from the number of atoms, its topological nature and its
resilience against control errors makes the proposal appealing for
experimental realization.

\section{Geometric phases}

Historically, the definition of geometric phase was originally introduced by
Berry in a context of closed, adiabatic, Schr\"odinger evolutions. What
Berry showed in his seminal paper~(Berry 1984) was that a quantum system
subjected to a slowly varying Hamiltonian manifests in its phase a geometric
behavior.

Let us summarize the derivation of the Berry phase. Consider a Hamiltonian
$H(\lambda)$ depending on some external parameters
$\lambda=(\lambda_1,\lambda_2,\dots, \lambda_m)$, and suppose that these
parameters can be varied arbitrarily inside a parameter space $\mathcal{M}$.
Assume that for each value of $\lambda$ the Hamiltonian has a completely
discrete spectrum of eigenvalues,
\begin{equation}\label{eq:spectr}
  H(\lambda)\ket{n(\lambda)}=E_n(\lambda)\ket{n(\lambda)},
\end{equation}
where $\ket{n(\lambda)}$ and $E_n(\lambda)$ are eigenstates and eigenvalues,
respectively, of $H(\lambda)$. Suppose that the values of $\lambda$ change
slowly, along a smooth path in $\mathcal{M}$. Under the adiabatic
approximation, a system initially prepared in an eigenstate
$|n(\lambda)\rangle$ it remains in the corresponding instantaneous
eigenspace.

In the simplest case of a non-degenerate eigenvalue, the evolution of the
eigenstate is specified  by the spectral decomposition~(\ref{eq:spectr}) up
to a phase factor. This phase factor can be evaluated by solving the
Schr\"odinger equation under the constraint of the adiabatic approximation,
yielding
\begin{equation}\label{esp:state}
\ket{\psi(t)_n}= e^{-i\beta}\exp\left\{i\oint_C\mathbf{A}\cdot
d\lambda\right\}
\ket{n(\lambda(t))},
\end{equation}
where $\beta=-i\int_0^T E_n(t) dt$ is the usual dynamical phase, and the
extra phase factor is the geometric phase. This phase has the form of a path
integral of a vector potential $\mathbf{A}$ (analogous to the
electromagnetical vector potential) called the Berry connection, whose
components are
\be
A_i=i\bra{n(\lambda)}\frac{\partial}{\partial
\lambda_i}\ket{n(\lambda)}.
\label{connection}
\ee
Berry was the first recognizing that this additional phase factor has an
inherent geometrical meaning: it cannot be expressed as a single valued
function of $\lambda$, but it is a function of the path followed by the
state during its evolution. Surprisingly, the value of this phase depends
only on the geometry of the path, and not on the rate at which it is
traversed. Hence the name ``geometric phase".

\begin{figure}
\begin{center}
\includegraphics[width=0.45\textwidth]{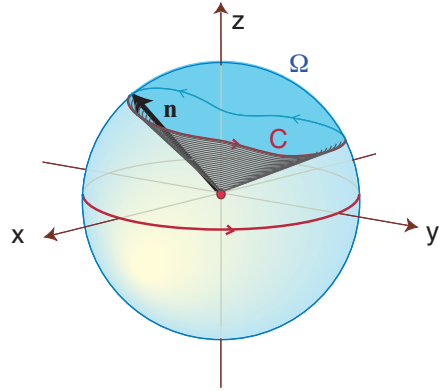}
\caption{The geometric phase is proportional to the solid angle
  spanned by the Hamiltonian with respect of its degeneracy point.}
\label{Bloch}
\end{center}
\end{figure}

The simplest, but still significant example of geometric phase, is the one
obtained for a two-level system, such as a spin-1/2 particle in the presence
of a magnetic field. Its Hamiltonian is given by
\be H(\theta,\phi)=\mathbf{B}(\theta,\phi)\cdot
\mbox{\boldmath{$\sigma$}} =|\mathbf{B}| U(\theta,\phi) \sigma^z
U^\dagger(\theta,\phi),
\ee
where $(\theta, \phi)$ determine the orientation of the magnetic field,
$U(\theta,\phi)=e^{-i\sigma^z\phi/2}e^{-i\sigma^y\theta/2}$ is a $SU(2)$
transformation which rotates the operator
$\mathbf{B}\cdot\mbox{\boldmath{$\sigma$}}$ to the $z$-direction and
$\mbox{\boldmath{$\sigma$}}=(\sigma^x,\sigma^y, \sigma^z)$ is  the vector of
Pauli's operators, given by
\be
\sigma^x =
\left(\begin{array}{cc}
0 & 1 \\
1 & 0
\end{array}\right) ,\Sp
\sigma^y =
\left(\begin{array}{cc}
0 & -i \\
i & 0
\end{array}\right) ,\Sp
\sigma^z =
\left(\begin{array}{cc}
1 & 0 \\
0 & -1
\end{array}\right).
\ee
With this parametrization the Hamiltonian can be represented as a vector on
a sphere, centered in the point of degeneracy of the Hamiltonian
($|\mathbf{B}|=0$), as seen in Figure~\ref{Bloch}.

For $\theta=\phi=0$ we have that $U=\one$ and the two eigenstates of the
system given by $|+\rangle=(1,0)^T$ and $|-\rangle=(0,1)^T$ with
corresponding eigenvalues $E_\pm=\pm|\mathbf{B}|/2$. Let us consider the
evolution resulting when a closed path $C$ is spanned adiabatically on the
sphere. Following the previous general consideration it is easy to show that
the Berry connection components corresponding, e.g. to the $|+\rangle$
state, are given by
\begin{equation}
A_\theta =0 ,\,\,\, A_{\phi}=\frac{1}{2}\left(1-\cos\theta \right)
\end{equation}
that leads to the geometric phase
\begin{equation}\label{spinphase}
\phi=\oint_{C}\mathbf{A}\cdot\\dr=\frac{1}{2}
\int_{\Sigma(\theta,\phi)}\!\!\!\sin{\theta}\;d\theta
d\phi=\frac{\Omega}{2}.
\end{equation}
The geometric phase that corresponds to the $|-\rangle$ state is given by
$\gamma(C)=-\Omega/2$. Here $\Omega=\int\!\!\int_{\Sigma}sin\theta d\theta d
\phi$ is the solid angle enclosed by the loop, as seen from the degeneracy
point. In this expression the geometric origin of the geometric phase $\phi$
is evident. Its value depends only on the way in which these parameters are
changed in relation with the degeneracy point of the Hamiltonian.

A particularly interesting case is the one in which the Hamiltonian can be
casted in a real form, corresponding to $\theta=\pi/2$. In this case the
phase becomes $\phi=\pi$ reproducing the sign change of the eigenstate, when
it circulates a point of degeneracy ($|\mathbf{B}|=0$), in agreement with
Longuet-Higgins theorem.

\section{The XY model and its criticality}

In order to illustrate the connection between geometric phases and critical
spin systems we shall consider the concrete example of a chain of spin-1/2
particles subject to XY interactions. This is a one dimensional model with
nearest neighbors spin-spin interactions, where we allow the presence of an
external magnetic field oriented along the $z$-direction. The Hamiltonian is
given by
\begin{equation}
\label{HXYModel}
H = -\sum_{l=-M}^{M} \left(\frac{1 + \gamma}{2}\sx_l
\sx_{l+1} + \frac{1 - \gamma}{2}\sy_l \sy_{l+1}
+\frac{\lambda}{2}\sz_l \right),\nonumber
\end{equation}
where $\sigma^\mu_l$ are the Pauli matrices at site $l$, $\gamma$ is the x-y
anisotropy parameter and $\lambda$  is the strength of the magnetic field.
This model was first explicitly solved by Lieb {\it et al.} (1961) and by
Katsura (1962). Since the XY model is exactly solvable and still presents a
rich structure it offers a benchmark to test the properties of geometric
phases in the proximity of criticalities.

In particular, we are interested in a generalization of Hamiltonian
(\ref{HXYModel}) obtained by applying to each spin a rotation with angle
$\phi$ around the $z$-direction
\begin{equation}
 \label{HXYphi}
H(\phi) = U(\phi)H U^\dag (\phi)\quad \text{ with }\quad U(\phi) =
\prod_{l=-M}^{M} e^{i\sz_l\phi/2},
\end{equation}
in the same way as we did for the single spin-1/2 particle. The family of
Hamiltonians generated by varying $\phi$ is clearly isospectral and,
therefore, has the same energy spectrum as the initial Hamiltonian. In
addition, due to the bilinear form of the interaction term we have that
$H(\phi)$ is $\pi$-periodic in $\phi$. The Hamiltonian $H(\phi)$ can be
diagonalized by a standard procedure based on the Jordan-Wigner
transformation and the Bogoliubov transformation (Carollo
\& Pachos 2005). From this procedure one can obtain the ground state,
$\ket{g}$, which is given by
\begin{equation}\label{groundstate}
\ket{g}\!=\!\!\!\prod_{\otimes k>0}\!\!\Big(\!\cos {\theta_k \over 2}
\ket{0}_{\!k}\ket{0}_{\!\!-k}
\!\!-ie^{2i\phi} \sin {\theta_k \over 2}
\ket{1}_{\!k} \ket{1}_{\!\!-k}\!\Big),
\end{equation}
where $\ket{0}_{k}$ and $\ket{1}_k$ are the vacuum and single
fermionic excitation of the k-th momentum mode. The angle $\theta_k$
is defined by $\cos\theta_k=\epsilon_k/\Lambda_k$ with
$\epsilon_k=\cos{2\pi k \over N}-\lambda$ and the energy gap above
the ground state is given by $\Lambda_k=\sqrt{\epsilon_k^2+ \gamma^2
\sin^2{2\pi k \over N}}$. It is remarkable that by inspection of
(\ref{groundstate}) the ground state can be interpreted as the direct
product of $N$ spins each one having its own orientation given by the
direction $(2\phi,\theta_k)$. As we will see in the next section this
observation will make the evaluation of the ground state geometric phase a
simple task. Before investigating the geometric properties of the XY model
we will first consider the behavior of the spectrum as a function of the
external parameters $\gamma$, $\lambda$ and $\phi$.

Let us first review the concept of quantum phase transitions. A many body
system, driven by a parameter $g$, undergoes a quantum phase transition at a
point $g=g_c$ when the energy density of the ground state at $g=g_c$ is
non-analytic. This point is associated with a crossing or an avoiding of the
energy eigenvalues. It is characterized by either a discontinuity in the
first derivative of the ground state energy density (first-order phase
transition) or by discontinuity or divergence in the second derivative of
the ground state energy density (second-order quantum phase transition)
assuming that the first derivative is continuous. In particular, the energy
gap $\Delta$ between the ground and the first excited states vanishes like
$\Delta\propto |g-g_c|^{z\nu}$ as $g$ approaches $g_c$ creating a point of
degeneracy that we will call critical. Moreover, the non-analyticity of the
energy eigenvalues is related to abrupt changes in the ground state
properties. As these transitions occur at zero temperature they are driven
purely by quantum fluctuations~(Sachdev 2001). It can be shown that the
length of the associated quantum correlations, $\xi$, diverges like
$\xi^{-1}\propto |g-g_c|^\nu$ as $g$ approaches the critical point $g_c$.
The parameters $z$ and $\nu$ are called the critical exponents and their
values are universal, independent of most of the microscopic details of the
system Hamiltonian.

For the case of the XY model one can identify the critical points by finding
the regions where the energy gap $\Lambda_k$ vanishes. Indeed, there are two
regions in the $\lambda$, $\gamma$ space that are critical. When $\gamma=0$
we have $\Lambda_k=0$ for $-1<\lambda<1$, which is a first order phase
transition with an actual energy crossing and critical exponents $z=2$ and
$\nu=1/2$. The other critical region is given by $\lambda=\pm 1$ where we
have $\Lambda_k=0$ for all $\gamma$. This is a second order quantum phase
transition with energy level avoiding. When $\gamma=1$ and $\lambda =\pm1$
we obtain the Ising critical model with critical exponents $z=1$ and
$\nu=1$.

\begin{figure}
\begin{center}
\includegraphics[width=4in]{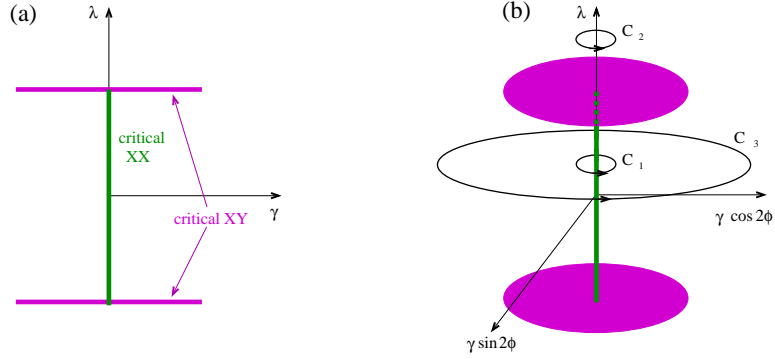}
\caption{(a) The regions of criticality of the XY Hamiltonian are
presented as a function of the parameters $ \lambda $ and $\gamma$ and (b)
the corresponding ones for the Hamiltonian $H(\phi)$ where $\phi$
parameterizes a rotation around the $\lambda$ axis. Possible paths for the
geometric evolutions are depicted spanned by varying the parameter $\phi$.}
\label{criticality}
\end{center}
\end{figure}

Finally, let us consider the criticality behavior of the rotated
Hamiltonian, $H(\phi)$. The energy gap, $\Lambda_k$ does not depend on the
angle $\phi$, as this parameter is related to an isospectral transformation.
Hence, the criticality region for the rotated Hamiltonian, $H(\phi)$, is
obtained just by a rotation of the critical points of the XY Hamiltonian
around the $\lambda$ axis. This is illustrated in Figure~\ref{criticality},
where the Ising type criticality corresponds now to two planes at $\lambda
=1 $ and $\lambda=-1$ and the XX criticality remains along the $\lambda$
axis for $|\lambda|<1$.

\section{Geometric phases and XY criticalities}

Figure 2 depicts the critical points of the XY model. Now we are interested
in spanning looping trajectories in the space comprising of the Hamiltonian
parameters $\lambda$, $\gamma$ and $\phi$. The aim is to determine the
geometric evolutions corresponding to these paths and relate them to the
criticality regions. A special family of paths is of interest that
circulates the $\lambda$ axis just by varying the $\phi$ parameter from zero
to $\pi$. Indeed, these paths enclose the XX criticality only when
$-1<\lambda<1$. As we shall see in the following it is possible to evaluate
the corresponding geometric phases of the ground and the first excited
states as a function of $\lambda$ and $\gamma$.

\begin{figure}
\begin{center}
\includegraphics[width=4.5in]{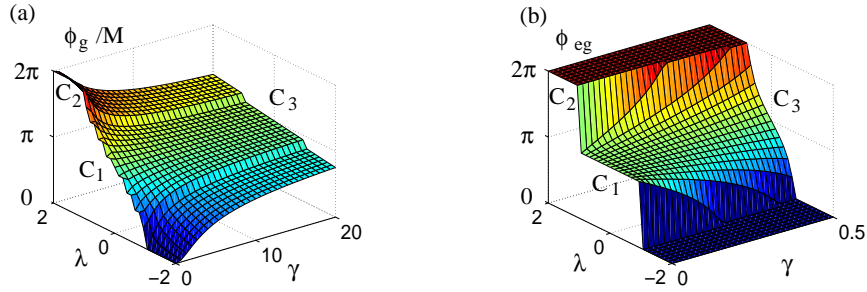}
\caption{The geometric phase corresponding to the ground state (a)
and the relative one between the ground and first excited state (b) as a
function of the path parameters $\lambda $ and $\gamma$. Each point of the
surface corresponds to the geometrical phase for a path that is spanned by
varying $\phi$ from $0$ to $\pi$ for certain $\lambda$ and $\gamma$. The
values of the geometric phase corresponding to the loops $C_1$, $C_2$ and
$C_3$ in Figure~\ref{criticality} are also indicated.}
\label{berry}
\end{center}
\end{figure}

Using the standard formula it is easy to show that the geometric phase of
the ground state $\ket{g}$ is given by
\begin{equation}
\label{geomphase}
\varphi_g = -i\int_0^\pi\bra{g}\frac{\partial}{\partial\phi}
\ket{g} = \sum_{k>0}\pi(1-\cos\theta_k).
\end{equation}
This result can be understood by considering the form of $\ket{g}$, which is
a tensor product of states, each lying in the two dimensional Hilbert space
spanned by $\ket{0}_{k}\ket{0}_{-k}$ and $\ket{1}_k\ket{1}_{-k}$. For each
value of $k(>0)$, the state in each of these two-dimensional Hilbert spaces
can be represented as a Bloch vector with coordinates $(2\phi,\theta_k)$. A
change in the parameter $\phi$ determines a rotation of each Bloch vector
about the $z$-direction. A closed circle will, therefore, produce an overall
phase given by the sum of the individual phases as given in
(\ref{geomphase}) and illustrated in Figure \ref{berry}(a).

\begin{figure}
\begin{center}
\includegraphics[width=3.3in]{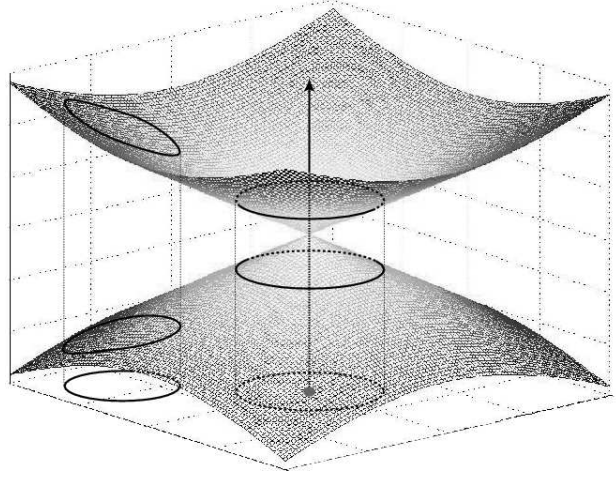}
\caption{The conical intersection between the two lowest energy levels
of the Hamiltonian as a function of its parameters. A contractible loop,
i.e. a loop that can be continuously deformed to a point of the domain,
produces a zero geometric phase. A non-trivial geometric phase is obtained
from non-contractible loops.}
\label{conical}
\end{center}
\end{figure}

Of particular interest is the relative geometric phase between the first
excited and ground states given by the difference of the geometric phases
acquired by these two states. The first excited state is given by
\begin{equation}\label{excitedstate}
\ket{e_{k_0}} =\!\!\!\ket{1}_{\!k_0}\ket{0}_{\!\!-k_0}
\prod_{\otimes k>0,\,\,k\neq \pm k_0}\!\!\Big(\!\cos {\theta_k \over 2}
\ket{0}_{\!k}\ket{0}_{\!\!-k} \!\!-ie^{2i\phi} \sin {\theta_k \over 2}
\ket{1}_{\!k} \ket{1}_{\!\!-k}\!\Big),
\end{equation}
with $k_0$ corresponding to the minimum value of the energy $\Lambda_k$. The
behavior of this state is similar to a direct product of only $N-1$ spins
oriented along $(2\phi,\theta_k)$ where the state of the spin corresponding
to momentum $k_0$ does not contribute any more to the geometric phase. Thus
the relative geometric phase between the ground and the excited states
becomes
\begin{equation}
\label{connectionExcited}
\varphi_{eg} \equiv \varphi_e-\varphi_g =
-\pi (1-\cos \theta_{k_0})
\end{equation}
In the thermodynamical limit ($N \to \infty$), $\phi_{eg}$ takes the form
\begin{equation}
\label{GPExcGrndSmallGamma}
\varphi_{eg}=\left\{\begin{array}{cl}
0,      &   \text{for $|\lambda |>1-\gamma^2$} \\
-\pi+{\pi \lambda \gamma \over
  \sqrt{(1-\gamma^2)(1-\gamma^2-\lambda^2)}},     &    \text{for
  $|\lambda |<1-\gamma^2$}
\end{array}\right.
\end{equation}
where the condition $|\lambda | > 1-\gamma^2$ constrains the excited state
to be completely oriented along the $z$-direction resulting in a zero
geometric phase. As can be seen from Figure~\ref{berry}(b), the most
interesting behavior of $\varphi_{eg}$ is obtained in the case of $\gamma$
small compared to $\lambda$. In this case $\varphi_{eg}$ behaves as a step
function, giving either $\pi$ or $0$ phase, depending on whether
$|\lambda|<1$ or $|\lambda|>1$, respectively. This behavior is precisely
determined from whether the corresponding loop encloses a critical point or
not and can be used as a witness of its presence. In particular, in the
$|\lambda |<1-\gamma^2$ case the first term corresponds to a purely
topological phase, while the second is a geometric contribution. Indeed, the
first part gives rise to a phase which depends solely on the topological
character of the trajectory traced by the $(2\phi,\theta_k)$ coordinates. In
particular if $n$ circulations are performed then the topological phase is
$n\pi$, where $n$ is the winding number. The second term is geometric in
nature and it can be made arbitrarily small by tuning appropriately the
couplings $\lambda$ or $\gamma$. This idea is illustrated in
Figure~\ref{conical}, where the energy surface of ground and first excited
state is depicted. The point of degeneracy is the intersection of the two
surfaces. This is the point where the energy density is not analytical.
Consider the case of a family of loops converging to a point. In the trivial
case where the limiting point does not coincide to a degeneracy, the
corresponding geometric phase converges to zero. If instead, the degeneracy
point is included, the geometric phase tends to a finite value~(Hamma 2006).

To better understand the properties of the relative geometric phase, we
focus on the region of parameters with $\gamma\ll 1$. In this case, it can
be shown (Carollo \& Pachos 2005) that the Hamiltonian, when restricted to
its lowest energy modes, can be casted in a {\em real} form and, for
$|\lambda|<1$, its eigenvalues present a {\em conical intersection} centered
at $\gamma=0$.
It is well known that when a closed path is taken within the real domain of
a Hamiltonian, a topological phase shift $\pi$ occurs only when a conical
intersection is enclosed. In the present case, the conical intersection
corresponds to a point of degeneracy where the XX criticality occurs and it
is revealed by the topological term in the relative geometric phase
$\varphi_{eg}$. It is worth noticing that the presence of a conical
intersection indicates that the energy gap scales linearly with respect to
the coupling $\gamma$ when approaching the degeneracy point. This implies
that the critical exponents of the energy, $z$, and of the correlation
length, $\nu$, satisfy the relation $z\nu=1$ which is indeed the case for
the XX criticality~(Sachdev 2001). In the following we shall see that
geometric phases are sufficient to determine the exact values of the
critical exponents and thus provide a complete characterization of the
criticality behavior.

\section{The general case}

We shall show here that the vacuum expectation value of a hermitian
operator, $O$, can be written in terms of a geometric phase. This is a
rather general result that can be used to study critical models, usually
probed by the behavior of vacuum expectation values of observables, just by
considering geometric phases. We assume, first, that $O$ does not commute
with the Hamiltonian, a requirement satisfied for the case of a
non-degenerate spectrum and, second, that $O$ can transform the ground state
in a cyclic fashion. The latter provides the looping trajectories of the
geometric evolutions.

To show that let us extend the initial Hamiltonian, $H_0$, of the model in
the following way
\be
H(\lambda) = H_0 +\lambda O
\ee
Turning to the interaction picture with respect to $O$ we obtain
\be
H_\text{int}(\lambda) = U(\lambda t)H_0 U^\dagger(\lambda t)
\ee
where $U(\lambda t) = \exp(-i \lambda O t)$. From the cyclicity requirement
there exists time $T$ such that the unitary rotation $U(\lambda T)$ takes
the ground state $\ket{\psi}$ to itself, i.e. $U(\lambda T)\ket{\psi}
=\ket{\psi}$. Hence, the desired cyclic evolution is obtained by a rotation
generated by $O$. The geometric phase that result from the cyclic evolution
is given by~(\ref{esp:state}) and, thus, we have
\begin{equation}
\varphi = \lambda T \bra{\psi} O \ket{\psi}
\label{Relation}
\end{equation}
Hence, the expectation value of an operator that can generate circulations
of the ground state is expressible with respect to a geometric phase.

One can easily verify this relation for the simple case of a spin-1/2
particle in a magnetic field. When the direction of the magnetic field is
changed adiabatically and isospectrally then the state of the spin is guided
in a cyclic path around the $z$-direction. The generated phase is given by
$\varphi = \pi(1- \cos \theta)$, where $\theta$ is the fixed direction of
the magnetic field with respect to the $z$-direction. On the other hand, one
can easily evaluate that the expectation value of the operator
$(1-\sigma^z)/2$ that generates the cyclic evolution is given by
$\bra{\psi}(1-\sigma^z)/2\ket{\psi} =(1-\cos\theta)/2$ which verifies
relation (\ref{Relation}) as for this example $\lambda T=2\pi$.

This connection has far reaching consequences. It is expected that intrinsic
properties of the state will be reflected in the properties of the geometric
phases. The latter, as they result from a physical evolution can be obtained
and measured in a conceptually straightforward way. Here, we are interested
in employing geometric phases to probe critical phenomena of spin systems.
Indeed, from the particular example of the XY model we saw that the presence
of critical points can be detected by the behavior of specific geometric
evolutions and the corresponding critical exponents can be extracted. This
comes as no surprise as one can choose geometric phases that correspond to
the classical correlations of the system (expectation values, e.g. of
$\sigma^z_1\sigma^z_L$) from where the correlation length and the critical
behavior can be obtained.

Let us apply this idea to the XY model we studied earlier. There the
rotations are generated by the operator $O= \sum_l \sigma^z_l$.
Hence, the resulting geometric phase is proportional to the total
magnetization
\be
M_z = \bra{\psi} \sum_l \sigma^z_l \ket{\psi}
\ee
It is well known (Sachdev 2001) that the mangetization $M_z$ can serve as an
order parameter, from which one can derive all the critical properties of
the XY model just by considering its scaling behavior. Indeed, Zhu 2006 has
considered the scaling of the ground state geometric phase of the XY model
from where he evaluated the Ising critical exponents. As it has been shown
here this is a general property that can be applied to any critical system.

\section{Physical implementation with optical lattices}

This construction, apart from its theoretical interest, offers a possible
experimental method to detect critical regions without the need to cross
them. When a physical system is forced to go through a critical region then
excited states may become populated due to the vanishing energy gap, thus
undermining the identification of the system state. Hence, being able to
probe the critical properties of a physical systems just by evolving it
around the critical area is of much interest to experimentalists as the
energy gap can be kept to a finite value.

In particular, we shall implement this model with optical lattices a system
that has proven versatile in the field of quantum simulations. To this end,
consider two bosonic species labelled by $\sigma=a, b$ that can be given by
two hyperfine levels of an atom. Each one can be trapped by a laser field
configured as a standing wave that is heavily detuned from any atomic
transitions. Thus, the atom acts as a dipole in the presence of a periodic
sinusoidal trapping potential that can generate one, two or three
dimensional lattices. Here we will restrict to the case where the atoms in
an arbitrary superposition of state $a$ and $b$ are confined in a
one dimensional array by the help of two in-phase optical lattices. The
tunnelling of atoms between neighboring sites is described by
\begin{equation}
V=-\sum_{l\sigma} (J_\sigma a_{l\sigma}^\dagger a_{(l+1)
\sigma} +\text{H.c.})
\end{equation}
where $a_\sigma$ and $a^\dagger_\sigma$ are the annihilation and creation
operators of particles $\sigma$ and $J_\sigma$ their corresponding
tunnelling coupling. When two or more atoms are present in the same site,
they experience collisions given by
\be
H^{(0)}=
\sum_{l\sigma \sigma'}  {U_{\sigma \sigma'} \over 2}
a^\dagger_{l\sigma} a^\dagger_{l\sigma'} a_{l\sigma'} a_{l\sigma}
\ee
where $U_{\sigma \sigma'}$ are the collisional couplings between atom
species $\sigma$ and $\sigma'$. We shall consider the limit $J \ll U$ where
the system is in the Mott insulator regime with one atom per lattice site
(Kastberg {\em et al.} 1995; Raithel {\it et al.} 1998). In this regime, the
effective evolution is obtained by adiabatic elimination of the states with
a population of two or more atoms per site, which are energetically
unfavorable. Hence, to describe the Hilbert space of interest, we can employ
the pseudospin basis of $|\!\!\uparrow\rangle\equiv |n_l^a=1,n_l^b=0\rangle$
and $|\!\!\downarrow\rangle\equiv |n_l^a=0,n_l^b=1\rangle $, for lattice
site $l$, and the effective evolution can be expressed in terms of the
corresponding Pauli (spin) operators.

It is easily verified by perturbation theory that when the tunnelling
coupling of both atomic species is activated, the following exchange
interaction is realized between neighboring sites~(Kuklov \& Svistunov 2003;
Duan {\it et al.} 2003)
\be
H_1=-\frac{J_a J_b}{U_{ab}} \sum_l \left(
\sigma^x_l\sigma^x_{l+1}+\sigma^y_l\sigma^y_{l+1}\right).
\label{Ham1}
\ee
In order to create an anisotropy between the $x$ and $y$ spin directions, we
activate a tunnelling by means of Raman couplings (Duan {\it et al.} 2003).
Application of two standing lasers $L_1$ and $L_2$, with zeros of their
intensities at the lattice sites and with phase difference $\phi$, can
induce tunnelling of the state $|+\rangle
\equiv ({e^{-i\phi /2} |a\rangle + e ^{i\phi /2} |b\rangle) /  \sqrt{2}}$.
The resulting tunnelling term is given by $V_c=J_c \sum_l (c^\dagger_l
c^{}_{l+1} +\text{H.c.})$, where $c_l$ is the annihilation operator of
$|+\rangle$ state particles. The tunnelling coupling, $J_c $, is given by
the potential barrier of the initial optical lattice superposed by the
potential reduction due to the Raman transition. The resulting evolution is
dominated by an effective Hamiltonian given, up to a readily compensated
Zeeman term, by
\be
H_2=-{1 \over 2}{J_c^2 \over U_{ab}} \sum_l
U(\phi)\sigma^x_l\sigma^x_{l+1}U^\dagger(\phi)
\label{Ham0}
\ee
where $U(\phi)$ was defined in (\ref{HXYphi}). Combining the rotationally
invariant Heisenberg interaction $H_1$ with $H_2$ gives the rotated XY
Hamiltonian described by equation~(\ref{HXYphi}), where the parameter
$\gamma$ is given by $J_c^2/(2\epsilon U_{ab})$ and $\epsilon = (2J_a J_b
+J_c^2/2)/U_{ab}$ is the overall energy scale multiplying the
Hamiltonian~(\ref{HXYModel}). The magnetic field term $\lambda\sum_l
\sigma^z_l$ is easily produced by a homogeneous and heavily detuned
laser radiation. If the radiation has amplitude $\Omega$ and detuning
$\Delta$ then the magnetic coupling is given by $\lambda=\Omega^2/\Delta$.
Finally, the angle $\phi$ of the rotated XY Hamiltonian is given by the
phase difference, $\phi$, of the lasers $L_1$ and $L_2$. Hence, the complete
control of the rotated XY Hamiltonian can be obtained by the optical lattice
configuration presented here and the geometric evolutions can be performed
by varying the phase $\phi$ from $0$ to $\pi$.

\section{Conclusions}

In this article, we have presented a method that theoretically, as well as
experimentally, allows for the detection of regions of criticality through
the geometric phase, without the need for the system to experience quantum
phase transitions. The latter is experimentally hard to realize as the
adiabaticity condition breaks down at the critical point and the state of
the system is no longer faithfully represented by the ground state.
Alternatively, the finite energy gap $\Lambda_k$ that is present along the
circular procedure is sufficient to adiabatically prevent unwanted
transitions between the ground state and the excited ones, even in the
thermodynamical limit.

The origin of the geometric phase can be ascribed to the existence of
degeneracy points in the parameter space of the Hamiltonian (Hamma 2006).
Hence, a criticality point can be detected by performing a looping
trajectory around it and detecting whether or not a non-zero geometric phase
has been generated. For the case of the XY model the topological nature of
the resulting phase pinned to the value, $\varphi_{eg}\approx \pi$, is
revealed by its resilience with respect to small deformations of the loop.
This characteristic results from the conical intersection structure of the
potential surfaces that is equivalent to having the critical exponents
satisfying $z\nu=1$. In addition, the critical exponents can be evaluated by
scaling arguments of the geometric phases (Zhu 2005). Hence, additional
information about the critical exponents can be deduced from the topological
nature and the exact value of the geometric phase. Moreover, topological
phases are inherently resilient against control errors, a property that can
be proved to be of a great advantage when considering many-body systems.
Such a study can be theoretically performed on any system which can be
analytically elaborated such as the case of the cluster Hamiltonian~(Pachos
\& Plenio 2004), or exploited numerically when analytic solutions are
not known. The generalization of these results to a wide variety of critical
phenomena and their relation to the critical exponents is a promising and
challenging question which deserves extensive future investigation.

\end{document}